\documentclass[twocolumn]{article}
\usepackage{graphics}
\begin{document}
\setlength{\unitlength}{1.0mm}
\title{Chaotic transients in the switching of roto-breathers}
\author{ Richard T Giles and Feodor V Kusmartsev 
  \\
Department of Physics, Loughborough University,
 Loughborough, LE11 3TU, U.K.}
\date{\today}
\maketitle

  By integrating a set of model equations for Josephson ladders
  subjected to a uniform transverse bias current we have found almost
  all of the kinds of breathers described in recent experiments, and
  closely reproduced their voltage-current characteristics and
  switching behaviour.  Our main result is that a chaotic transient
  occurs in the switching process. The growth of tiny perturbations
  during the chaotic transient causes the new breather configuration
  to be extremely sensitive to the precise history of the initial
  breather and can also cause the new breather to have a new centre of
  symmetry.

\medskip
\hrule
\bigskip

\begin{figure}[htp]
\begin{center}
 \resizebox{60mm}{!}
   {\includegraphics*[50mm,50mm][170mm,210mm]{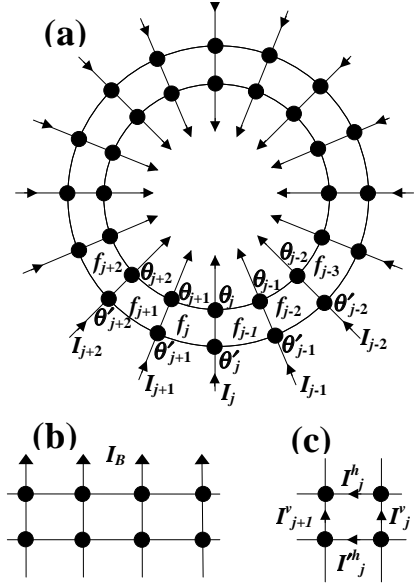}}
 \end{center}
\caption{\label{fig:ladders}(a) An annular ladder subjected to
  transverse bias currents~$I_j$. Each circle represents a
  superconducting island. Each link between islands represents a
  Josephson junction. $\theta$ is the phase of the superconducting
  order parameter and $f$ is the flux threading a plaquette. (b) A
  linear ladder. (c) Explanation of the notation used in
  Eq.~(\ref{eq:induction}).}
\end{figure}

The recent discovery\cite{binder00,trias00,binder00a} of
roto-breathers in Josephson ladder arrays (Fig.~\ref{fig:ladders})
driven by a transverse dc bias current have shown not only that these
localised excitations exist but that they exhibit remarkable
behaviour.

The Josephson ladder has been studied theoretically for many years. In
the absence of a driving current the interaction between vortices has
been found to be exponential and this leads to the vortex density
exhibiting a devil's staircase as the magnetic field is
increased\cite{kardar84,giles99}. Quantum fluctuations\cite{kardar86},
meta-stable states\cite{mazo95} and inductance effects\cite{mazo96}
have been studied as has the interaction of
vortices with a transverse dc bias current\cite{kim97a,barahona98}.

\begin{figure}[htp]
\begin{center}
 \resizebox{70mm}{!}
  {\includegraphics{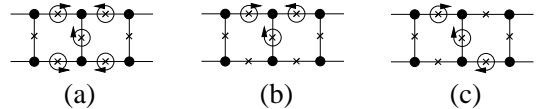}}
\end{center}
\caption{\label{fig:configs}Three configurations for the rotation of
  horizontal junctions around a breather with $N_R=1$ vertical
  rotating junctions: (a) the \emph{I}-breather, (b) the
  \emph{T}-breather, and (c) the \emph{Z}-breather.}
\end{figure}

A roto-breather\cite{takeno96} is a stable group of vertical junctions rotating
together ($\theta_j^\prime-\theta_j\approx\omega t$, see
Fig.~\ref{fig:ladders}). Such solutions have been studied
theoretically\cite{martinez99} in ladders driven by a uniform
transverse \emph{ac} bias current and are also stable\cite{mackay98} for the case of a \emph{dc} bias current. Rotation at a single vertical junction was
studied numerically\cite{mazo99,flach99} and two experimental
groups\cite{binder00,trias00} have now independently confirmed the
existence of such solutions and discovered further unpredicted
behaviour. In an annular ladder\cite{binder00}
(Fig.~\ref{fig:ladders}a) and in a linear
ladder\cite{trias00,binder00a} (Fig.~\ref{fig:ladders}b) breathers
were observed with various numbers of ``vertical'' (or radial)
junctions rotating. The voltage drop associated with rotating vertical
junctions causes some of the horizontal junctions to rotate and
various configurations were commonly observed
(Fig.~\ref{fig:configs}).

However the most interesting and puzzling feature was the switching
between breathers as the bias current was varied. Once a breather has
been initialised it is stable, but if the bias current is slowly
decreased then, at some critical value of the current, the
configuration of rotating junctions may change thus forming a new
breather.  Furthermore, it appears from the data that \textit{the new
  breather may have a different centre of symmetry from the old
  breather despite the symmetry of the ladder itself and the symmetry
  of the driving currents about a single vertical junction}.  We
present here the first numerical studies of the switching behaviour of
roto-breathers and show that this effect arises from the exponential
growth of even infinitesimal perturbations during a chaotic transient
which is associated with the switching process.

Aiming to describe the experiments as accurately as possible, we first
built up a model in which the current~$I$ through a junction is
determined by the well-established RSJ model equation
\begin{equation}
\frac{I}{I_{c}} = \frac{d^2\varphi}{dt^2}+\alpha\frac{d\varphi}{dt}+
\sin \varphi
\label{eq:rsj}
\end{equation}
where $I_{c}$ is the critical current and
$\varphi=\Delta\theta-\frac{2\pi}{\Phi_0}\int\mathbf{A}.\mathrm{d}\mathbf{l}$
where $\Delta\theta$ is the change in superconducting order
parameter~$\theta$ across the junction and $\mathbf{A}$ is the vector
potential.  The ``vertical'' (i.e.  radial) junctions may differ in
area from the ``horizontal'' junctions by an anisotropy parameter
$\eta=I_{ch}/I_{cv}=C_h/C_v=R_v/R_h$ where $I_{ch}$ ($I_{cv}$), $C_h$
($C_v$) and $R_h$ ($R_v$) are, respectively, the critical current,
capacitance and resistance of a horizontal (vertical) junction.  From
Fig.~\ref{fig:ladders}a we see that there are three unknowns per
plaquette: $\theta_j$, $\theta_j^\prime$ and $f_j$. To solve for these
unknowns we construct three equations per plaquette as follows. The
first two equations are obtained from current conservation at the top
(inner) and bottom (outer) rails. The third equation is obtained by
making the approximation that the induced flux $f_j-f_a$ (where $f_a$
is the applied flux, assumed constant) is produced solely by the
currents flowing around the immediate perimeter of the $j$th plaquette
(Fig.~\ref{fig:ladders}c):
\begin{equation}
\frac{f_j-f_a}{\Phi_0} =\frac{\beta_L}{8\pi}
\left(I_{j+1}^v-I_j^{h}-I_{j}^v+I_j^{\prime h}\right)
\label{eq:induction}
\end{equation}
where $\beta_L$ is an inductance parameter and we have chosen
$I_{cv}=1$.  Eq.~(\ref{eq:induction}) introduces a significant
difference from previous models\cite{mazo99,flach99} which assume that
the induced field is proportional to the loop (or ``mesh'') current
circulating the plaquette.  For square plaquettes,
Eq.~(\ref{eq:induction}) represents an improvement since it properly
takes into account the flux produced by the vertical conductors, although
it is still not a full nearest-neighbour mutual inductance
calculation\cite{dominguez96}.

Using Landau gauge we then have the following three coupled
differential equations for each plaquette:
\begin{eqnarray}
    \label{eq:-}
\lefteqn{\left\{\frac{d}{dt^2}+\alpha\frac{d}{dt}\right\}\left(-\theta_{j-1}^-
+4\theta_j^- -\theta_{j+1}^-\right)=2I_j+
 \sin\theta_{j-1}^-} \nonumber \ \ \ \ \ \ \ \ \ \\
& &  -4\sin\theta_j^- +\sin\theta_{j+1}^-
     + \frac{8\pi}{\beta_L}(f_{j-1}-f_j)
\end{eqnarray}
\begin{eqnarray}
\label{eq:+}
\lefteqn{\left\{\frac{d}{dt^2}+\alpha\frac{d}{dt}\right\}
\left(\chi_j^+-\chi_{j-1}^+\right)=} \nonumber \ \ \ \ \ \ \ \ \ \ \ \ \ \ \\
& & \sin\chi_{j-1}^+\cos\chi_{j-1}^- - \sin\chi_{j}^+\cos\chi_{j}^-
\end{eqnarray}
\begin{eqnarray}
\lefteqn{\left\{\frac{d}{dt^2}+\alpha\frac{d}{dt}\right\}
\left(\theta_{j+1}^- -\theta_j^-+2\eta\chi_j^-\right) =}
\nonumber \ \ \ \ \ \ \ \ \ \ \ \\ 
& & \sin\theta_j^- - \sin\theta_{j+1}^-+
\frac{8\pi}{\beta_L}(f_{j}-f_a) -2\eta\sin\chi_j^-
\label{eq:f}
\end{eqnarray}
where $\theta_j^-=\theta_j^\prime-\theta_j$,
$\theta_j^+=\theta_j^\prime+\theta_j$,
$\chi_j^-=\frac{1}{2}(\theta^-_{j}-\theta^-_{j+1}-2\pi f_j)$,
$\chi_j^+=\frac{1}{2}(\theta^+_{j}-\theta^+_{j+1}-2\pi F/N)$ and $N$
is the number of plaquettes. Putting $F=0$, this is a sufficient set
of equations for describing the linear ladder (with the imposition of
the appropriate boundary conditions). For the case of the annular
ladder we also need one further equation for the loop flux $F$
(proportional to the total current circulating around the annulus):
\begin{equation}
\label{eq:F}
\frac{d^2F}{dt^2}+\alpha\frac{dF}{dt}=\frac{N^2(F_a-F)}{2\eta B_L} +
\frac{N}{2\pi}\sum_{j=1}^N\sin\chi_j^+ \cos\chi_j^-
\end{equation}
where $F_a$ is the applied flux, assumed constant, and $B_L$ is the
annulus inductance parameter. Note that the equations remain invariant
if all $\theta^+_j$ are rotated by the same amount, an amount which
can also be varied with time. It is only differences in $\theta^+$
from one site to another which are important. Note also that the
equations remain invariant if $\theta_j^-$ and/or $\theta_j^+$ are
rotated by $4\pi$ (\textit{not} $2\pi$).

In the RSJ model, on which our equations are based, each ideal
Josephson junction has in parallel with it a capacitance and a
resistance, or in other words a frequency filter.  The cut-off
frequency is $1/(RC)=\alpha$. Thus voltage oscillations of frequency
$\omega\gg\alpha$ tend to be suppressed, being shorted out by the
capacitance. In the experiments, and in our simulations, the rotating
junctions have $d\theta/dt \gg \alpha$ and so we can expect voltage
oscillations to be suppressed by a large factor. We might therefore
expect that all observed attractors should be characterised by having
all rates of rotation, currents and fluxes at more or less fixed
values. Many junctions will not rotate at all, in which case
Eq.(\ref{eq:rsj}) gives us $I=I_c\sin\phi$. For those that rotate
Eq.(\ref{eq:rsj}) gives us $I=\alpha d\phi/dt$, i.e. the supercurrent
and capacitative current more or less cancel one another. We call this
the \textbf{\emph{dc approximation}}.  Using this approximation one
can solve for the distribution of currents, voltages and fluxes
for any configuration of rotating junctions.  The
observed roto-breathers are in fact quite well described by this
approximation.

We now focus on determining whether or not our model exhibits the
interesting switching and symmetry breaking behaviour observed in the
data of Binder $et$ $al$\cite{binder00,binder00a}. Attempting to mimic
the experiments, we use a random number generator to give $\theta_j$,
$\theta_j^\prime$ and $f_j$ extremely small, but non-zero, initial
values at $t=0$. If instead, the initial state were taken as having
all $\theta_j=\theta_j^\prime=0$ then we could neglect Eq.(\ref{eq:+})
and all solutions would maintain the symmetry $\chi^+_{j-1}=\chi^+_j$
for all $j$; however it turns out\cite{mazo99} that this symmetric
state can be unstable to small perturbations.  Let
\begin{equation}
I_j=\left\{ \begin{array}{ll}
    I_B + I_\Delta & \mbox{ if $j=0$} \\
    I_B            & \mbox{ otherwise}
    \end{array}
\right.
\end{equation}
where $I_B$ is called the bias current.  Again following the
experiment, we slowly increase $I_\Delta$ while keeping $I_B=0$ until
rotation starts at site~$j=0$. $I_\Delta$ is then slowly decreased
while at the same time increasing $I_B$ to keep $I_0$ constant. When
$I_B$ has reached the desired value it is then held fixed while
$I_\Delta$ is slowly reduced to zero. At this point, provided $I_B$ is
not too large or small, we have just one rotating vertical junction
(at $j=0$), the horizontal junctions rotating in the
\emph{I}-configuration (Fig.~\ref{fig:configs}). We call this an
\emph{I}-breather with $N_R=1$ rotating vertical junctions. In
Fig.~\ref{fig:configs}a the \emph{I}-breather appears to have top-bottom
symmetry, but in fact this symmetry is broken since $\theta^+$ (which
is more or less static) actually varies from site to site.  Also the
currents circulating the two plaquettes on either side of the rotating
junction cause a static flux pattern rather like a
vortex-antivortex pair, the value of the flux in each plaquette being
what one would deduce in the dc approximation.  This must cause a
compressive force on the breather and an attractive force with the end
of any nearby breather of the same kind.

\begin{figure}[htp]
\begin{center}
   \includegraphics[80pt,460pt][280pt,670pt]{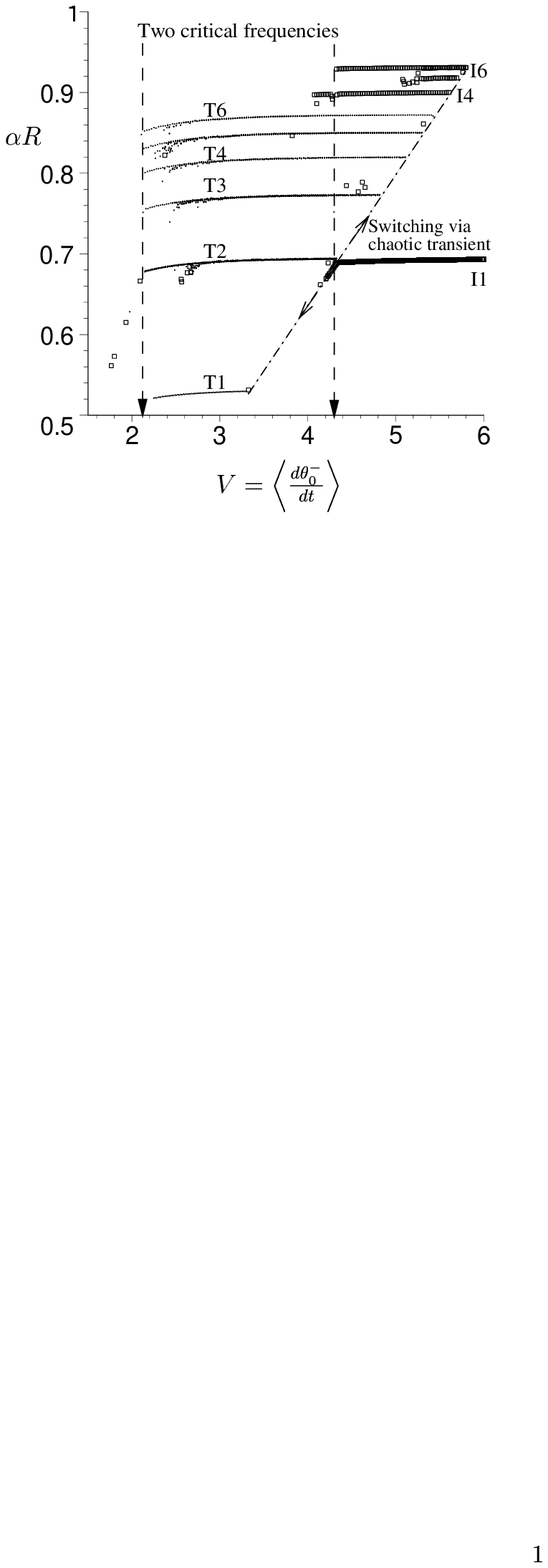}
\end{center}
\caption{\label{fig:vi}Resistance~$\alpha R$ (where $R=V/I_B$) plotted against
  voltage~$V=\left\langle d\theta_0^-/dt\right\rangle $. The results
  were obtained for an annular ladder with $N=8$ vertical junctions,
  $\alpha=0.07$, $\eta=0.44$, $\beta_L=4.8$, $B_L=F_a=f_a=0$. The
  middle vertical junction was excited in the manner described in the
  text and then the uniform bias current~$I_B$ was ramped down. The
  data come from many runs with different starting values of $I_B$.
  The labels T2, I1 etc. refer to breather types, e.g. ``I1'' means
  the one-site \emph{I}-breather.}
\end{figure}

Finally $I_B$ is slowly reduced while keeping $I_\Delta=0$. The
voltage current characteristics of the breathers produced as $I_B$ is
ramped down are shown in Fig.~\ref{fig:vi}. We show here only the
results for an annular ladder with $N=8$ vertical junctions and the
parameter values otherwise chosen to mimic the annular ladder
experiment\cite{binder00,private}. We have also put $B_L=0$ since we find that
although the inner region of the ladder tends to pick up trapped flux
(a vortex) in the breather switching process, this has little
influence on the voltage-current characteristics for this range of
parameter values.  Note that while \emph{I} and \emph{T}-breathers
occur, there are no \emph{Z}-breathers.  This is because in the
annular ladder the two ends of a breather are connected by a quiescent
region in which all superconducting order parameters must be static or
rotate together in synchrony (apart from small vibrations), and so
\emph{Z}-breathers cannot exist in the absence of other breathers.

All of the breathers are well characterised by a constant
resistance~$R$ (i.e. $V=\left\langle d\theta^-/dt\right\rangle =I_B
R$) and the value of $R$ agrees very closely with the value expected
in the dc approximation. Furthermore we see that at a critical
rotation frequency the single-site ($N_R=1$) \emph{I}-breather
generally broadens into a multi-site \emph{I}-breather, but may also
convert to a single-site or multi-site \emph{T}-breather. In agreement
with the experiments, this
occurs when the rotation rate of the horizontal junctions is close to
the minimum rotation rate that could be supported if the junctions
were isolated and driven by a constant current (cf. the miminum
rotation rate of an underdamped pendulum driven by a constant torque).
This critical frequency is more or less independent of $\alpha$ (for small
$\alpha$).  The multi-site \emph{I}-breathers switch to
\emph{T}-breathers at the same critical frequency. In the switching,
two of the four rotating horizontal junctions stop while the other two
double in frequency (to maintain Kirchoff's Law). As $I_B$ is reduced
further the rotation rate of the junctions falls again until the same
critical rotation rate is reached at which point all motion ceases. We
obtain qualitatively similar results both for longer (and shorter)
ladders and for linear ladders with similar parameter values.  The
main difference in linear ladders is that \emph{Z}-breathers are
produced also.  These observations agree very well with the behaviour
reported for a linear ladder with similar parameter
values\cite{binder00a}.  The annular ladder experiment\cite{binder00}
did not find any \emph{T}-breathers but his may very well be due to a
construction fault\cite{annular}.

Ref.~\cite{binder00a} shows that the linear ladder can support at
least several types of breathers. Nearly all of the configurations
found experimentally arose naturally in our simulations.  Furthermore
we found that in a multi-site breather, it appears that any two
adjacent vertical junctions can rotate in phase or $180^\circ$ out of
phase. It is clear that a great many breather configurations are
possible.

The main result of the simulations relates to the
switching to multi-site breathers. Although the single-site
\emph{I}-breather possesses left-right symmetry
about $j=0$, the multi-site breather produced in the switching process
frequently has a different centre of symmetry. Any
breather with an even number~$N_R$ of rotating junctions cannot have
its centre of symmetry at $j=0$.  Such breathers are produced even in
the annular ladder (where translational invariance is exact) and even
if we set all variables to zero initially, in which case the dynamical
equations, initial conditions and injected currents all have \textit{exact
  left-right symmetry} about site $j=0$.
The origin of the symmetry breaking lies in the occurrence
of chaotic dynamics during the switching.  Further
evidence of chaotic behaviour is provided by the fact that even the
smallest change in initial conditions or of any parameter usually
results in the production of a completely different kind of multi-site
breather. Both of these effects (i.e. symmetry breaking and extreme
sensitivity) were noticed in the experiments\cite{private}.

\begin{figure}[htb]
\begin{center}
  \includegraphics[78pt,554pt][297pt,655pt]{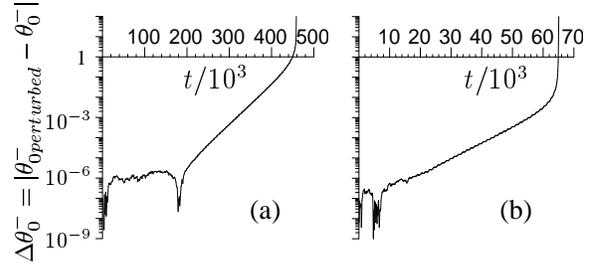}
\end{center}
\caption{\label{fig:lya}Exponential growth of perturbations during the chaotic
  transient which occurs when the single-site \emph{I}-breather
  becomes unstable at $V=\left\langle d\theta^-/dt\right\rangle =4.3$.
  (a) The initial state ($t=0$) was prepared with $I_B=0.45235$.  A
  perturbed initial state was then produced by introducing small
  random changes to all state variables.  After an initial period
  $\Delta\theta_0^-$ starts to grow exponentially with time. Switching
  to a new stable breather takes place at $t=460,000$ and produces the
  sudden rise in $\Delta\theta_0^-$ at that time. (b) Same except the
  initial state was prepared with $I_B=0.45234$.}
\end{figure}

To study the origin of these effects we have looked more closely at
what happens when the single-site \emph{I}-breather becomes unstable
at $V=4.3$ and switching takes place (see Fig.~\ref{fig:vi}).
Approaching the switching instability by reducing~$I_B$ very slowly, we find
that at $I_B=0.4524$ the breather appears to be stable and periodic
with no positive Lyapunov exponent.  However when the current is
lowered to $I_B=0.45235$ the motion becomes weakly chaotic with a
maximum Lyapunov exponent (Fig.~\ref{fig:lya}a) of $5\times10^{-5}$;
but the chaotic motion is itself unstable and switching to a new
stable breather eventually takes place after $460\,000$ time units.
Thus \emph{the switching between breathers occurs via a chaotic
  transient\cite{hilborn94}}. Clearly we have here the origin of the observed symmetry
breaking and extreme sensitivity.  The nature of the resulting final
breather is extremely sensitive to the smallest change in the
simulation (as one might expect given the chaotic nature of the
motion) but the time before switching occurs is more or less fixed.
If, before waiting for switching to occur, the bias current is slowly
lowered further to 0.45234 then the maximum Lyapunov exponent
increases to $2\times 10^{-4}$ and switching occurs after only
$65\,000$ time units (Fig.~\ref{fig:lya}b).

The only clear failures in the predictive power of the model occurred
near the superconducting gap frequency where the RSJ model is bound to
fail. In this region the experiments found that breather resistances
changed and that \emph{T}-breathers could be converted to
\emph{I}-breathers; neither of these effects were reproduced by the
simulations.

We conclude that our simulations exhibit most of the main features of
the roto-breathers recently observed in Josephson ladder arrays. In
addition to reproducing the experiments we have found that switching
occurs via a chaotic transient, and this may be a general feature of
the switching of roto-breathers in Josephson arrays. This chaotic
transient has some practical consequences. First of all it leads to
the extreme sensitivity and symmetry breaking already observed in the
experiments.  Secondly, since it has has a very broad Fourier
spectrum, characteristic of chaotic motion, we can predict the
existence of a source of noise associated with switching. The duration
of the noise (i.e. the transient) should depend on how quickly the
bias current is ramped (Fig.~\ref{fig:lya}). We hope that future
experiments will discover the noise and confirm this effect.

\bigskip

We are grateful to A.V.~Ustinov and P.~Binder for kindly
giving us their data and explaining it prior to publication.  We are
also grateful for discussions with M.V.~Fistul, S.~Flach, A.~Osbaldestin
and J.H.~Samson and for the hospitality of the Max-Planck-Institut,
Dresden and the Universit\"{a}t Erlangen-N\"{u}rnberg.


\begin{thebibliography}{10}

\bibitem{binder00}
P. Binder {\it et~al.}, \textit{Phys. Rev. Lett.} {\bf 84},  745  (2000).

\bibitem{trias00}
E. Tr{\'{\i}}as, J. Mazo, and T. Orlando, \textit{Phys. Rev. Lett.} {\bf 84},
  741  (2000).

\bibitem{binder00a}
P. Binder, D. Abraimov, and A. Ustinov, \textit{Phys. Rev. E} {\bf 62},  2858
  (2000).

\bibitem{kardar84}
M. Kardar, \textit{Phys. Rev. B} {\bf 30},  6368  (1984).

\bibitem{giles99}
R. Giles and F. Kusmartsev, \textit{J. Low Temp. Phys.} {\bf 117},  623
  (1999).

\bibitem{kardar86}
M. Kardar, \textit{Phys. Rev. B} {\bf 33},  3125  (1986).

\bibitem{mazo95}
J. Mazo, F. Falo, and L. Flor{\'{\i}}a, \textit{Phys. Rev. B} {\bf 52},  10433
  (1995).

\bibitem{mazo96}
J. Mazo and J. Ciria, \textit{Phys. Rev. B} {\bf 54},  16068  (1996).

\bibitem{kim97a}
S. Kim, \textit{Phys. Lett. A} {\bf 235},  408  (1997).

\bibitem{barahona98}
M. Barahona, S. Strogatz, and T. Orlando, \textit{Phys. Rev. B} {\bf 57},  1181
   (1998), and references therein.

\bibitem{takeno96}
S. Takeno and M. Peyrard, \textit{Physica D} {\bf 92},  140  (1996).

\bibitem{martinez99}
P. Martinez, L. Flor{\'{\i}}a, F. Falo, and J. Mazo, \textit{Europhys. Lett.}
  {\bf 45},  444  (1999).

\bibitem{mackay98}
R. MacKay and J.-A. Sepulchre, \textit{Physica D} {\bf 119},  148  (1998).

\bibitem{mazo99}
J. Mazo, E. Tr{\'{\i}}as, and T. Orlando, \textit{Phys. Rev. B} {\bf 59},
  13604  (1999).

\bibitem{flach99}
S. Flach and M. Spicci, \textit{J. Phys.: Condens. Matter} {\bf 11},  321
  (1999).

\bibitem{dominguez96}
D. Dom{\'{\i}}nguez and J. Jos{\'{e}}, \textit{Phys. Rev. B} {\bf 53},  11692
  (1996).

\bibitem{private}
P. Binder, private communication.

\bibitem{annular}
See footnote 14 of Ref.~\cite{binder00a}.

\bibitem{hilborn94}
R. Hilborn, {\em Chaos and Nonlinear Dynamics} (Oxford, UK, 1994).

\end{thebibliography}

\end{document}